\documentclass[12pt]{article}
\usepackage{myart}

\normalbaselines
\input tcilatex.tex

\begin{document}

\author{Yuri A. Rylov}
\title{Hydrodynamical interpretation of quantum mechanics: the momentum
distribution.}
\date{Institute for Problems in Mechanics, Russian Academy of Sciences\\
101, bld. 1, Vernadski Ave. Moscow, 119526, Russia \\
email: rylov@ipmnet.ru\\
Web site: {$http://rsfq1.physics.sunysb.edu/\symbol{126}rylov/yrylov.htm$}\\
}
\maketitle

\begin{abstract}
The quantum mechanics is considered to be a partial case of the stochastic
system dynamics. It is shown that the wave function describes the state of
statistically averaged system $\left\langle \mathcal{S}_{\mathrm{st}%
}\right\rangle $, but not that of the individual stochastic system $\mathcal{%
S}_{\mathrm{st}}$. It is a common practice to think that such a construction
of quantum mechanics contains hidden variables, and it is incompatible with
the von Neumann's theorem on hidden variables. It is shown that the original
conditions of the von Neumann's theorem are not satisfied. In particular,
the quantum mechanics cannot describe the particle momentum distribution.
The distribution $w\left( \mathbf{p}\right) =\left| \psi _{\mathbf{p}%
}\right| ^{2}$ is not a particle momentum distribution at the state $\psi $,
because it cannot be attributed to a wave function. It is closer to the mean
momentum distribution, although the two distributions do not coincide
exactly.
\end{abstract}

\newpage

\section{Introduction}

In the present paper we consider the quantum mechanics as a special case of
the stochastic systems dynamics \cite{R002c}. The dynamics of stochastic
systems describes any stochastic systems, but not only the quantum ones, and
dynamics of stochastic systems cannot be founded on the principles of
quantum mechanics, which are specific only for description of the
nonrelativistic quantum phenomena. In the stochastic system dynamics the
quantum principles appear only on the dynamical level as \textit{dynamic
properties }of quantum systems. The stochastic system dynamics is a \textit{%
usual statistical conception}, which contains two sorts of objects: (1) the
individual stochastic system $\mathcal{S}_{\mathrm{st}}$ and (2) the
statistical average system $\left\langle \mathcal{S}_{\mathrm{st}%
}\right\rangle $, which is the statistical ensemble $\mathcal{E}\left[ 
\mathcal{S}_{\mathrm{st}}\right] $ of stochastic systems $\mathcal{S}_{%
\mathrm{st}}$ normalized to one system. Dynamic equations describe the state
evolution of $\left\langle \mathcal{S}_{\mathrm{st}}\right\rangle $ and have
no relation to $\mathcal{S}_{\mathrm{st}}$.

The conventional quantum mechanics is also a statistical conception, but it
is an exotic statistical conception, because it contains only one sort of
objects: the quantum system $\mathcal{S}_{\mathrm{q}}$. What is the quantum
system $\mathcal{S}_{\mathrm{q}}$? Is it $\mathcal{S}_{\mathrm{st}}$, or $%
\left\langle \mathcal{S}_{\mathrm{st}}\right\rangle $? Different authors
answer this question differently. The conventional (Copenhagen)
interpretation of quantum mechanics supposes that $\mathcal{S}_{\mathrm{q}}$
is an individual system (i.e. $\mathcal{S}_{\mathrm{st}}$). But all authors
consider that the quantum mechanical conception contains only one sort of
objects: $\mathcal{S}_{\mathrm{q}}$. Dynamic equations describe the state
evolution of $\mathcal{S}_{\mathrm{q}}$.

Model conception of quantum phenomena (MCQP) is the name of the quantum
mechanics considered to be a special case of the stochastic systems
dynamics. The conventional quantum mechanics will be referred to as the
axiomatic conception of quantum phenomena (ACQP). These abbreviations will
be used for brevity.

In MCQP the hydrodynamic interpretation of quantum mechanics is considered
to be a primary interpretation, and the conventional interpretation in terms
of the wave function appears to be a secondary (derivative) interpretation.
Such an approach has a logical foundation, and we shall present it. Besides,
MCQP and the hydrodynamic interpretation of quantum mechanics can be easily
expanded onto the relativistic quantum phenomena, because MCQP takes into
account the quantum specific only on the dynamical level. To describe the
relativistic phenomena, it is sufficient to consider relativistic Lagrangian
of $\left\langle \mathcal{S}_{\mathrm{st}}\right\rangle $.

In ACQP the quantum specific is taken into account on the conceptual level.
The wave function (the main object of ACQP) is nonrelativistic, and
transition to the relativistic phenomena description is connected with a
revision of the conceptual structure of ACQP. Attempts of unification of the
nonrelativistic QM technique with the relativity principles failed. At any
rate, the expansion of conventional interpretation in terms of the wave
function meets with difficulties. As a result of these difficulties the
collision of relativistic particles is described only in terms of the $S$%
-matrix, because the detailed description of the collision process in terms
of the wave function appears to be impossible. In particular, one cannot
describe in terms of the wave function the details of the pair production
mechanism \cite{R003}.

Note that MCQP may be considered to be a theory with hidden variables. It is
believed that the quantum theory with hidden variables is impossible,
because according to the von Neumann's theorem on hidden variables \cite{N32}
(chp.4, sec. 1,2) it is incompatible with the conventional conception of
quantum mechanics. The mathematical formalism of ACQP and, in particular,
the theorem of von Neumann on hidden variables are founded on the following
statement.

For any observable quantity $\mathcal{R}$ and for any state $\varphi $ of
the considered quantum system the relation 
\begin{equation}
\mathrm{Erv}\left( \mathcal{R},\varphi \right) =\left( R\varphi ,\varphi
\right)  \label{h0.1}
\end{equation}
takes place. Here $\mathrm{Erv}\left( \mathcal{R},\varphi \right) $ is a
mathematical expectation of the quantity $\mathcal{R}$ at the state $\varphi 
$. The quantity\ $R$ is the operator of the observable quantity $\mathcal{R}$%
, and $\left( R\varphi ,\varphi \right) $ is the scalar product of two
vectors $R\varphi $ and $\varphi $ of the Hilbert space. It is supposed that
any observable quantity $\mathcal{R}$ can be measured instantaneously and
attributed to the state (wave function) $\varphi $ at this moment. The
formula (\ref{h0.1}) is supposed to be valid \textit{for all observable
quantities} $\mathcal{R}$ \textit{and for all states} $\varphi $ (wave
functions). If the formula (\ref{h0.1}) is valid not for all quantities $%
\mathcal{R}$, the mathematical formalism of ACQP cannot be founded, and the
theorem on hidden variables appears to be unproved.

We are going to show that the formula (\ref{h0.1}) is not valid in the case,
when $R=F\left( \mathbf{p}\right) $, where $F$ is an arbitrary function and $%
\mathbf{p}=-i\hbar \mathbf{\nabla }$ is the momentum operator. In
particular, the measurement of the momentum of a free particle needs a long
time. During this time the wave function $\psi $ of the particle changes
essentially, and the measured value $\mathbf{p}^{\prime }$ of the momentum
cannot be attributed to any definite state (wave function). As a result the
formula (\ref{h0.1}) is not satisfied for such a quantity as the particle
momentum.

Indeed, the wave function for the free particle of the mass $m$ has the form 
\begin{equation}
\psi \left( t,\mathbf{x}\right) =\frac{1}{\left( 2\pi \hbar \right) ^{3/2}}%
\int \left| \psi _{\mathbf{p}}\right| e^{-i\zeta \left( \mathbf{p}\right)
}\exp \left( i\frac{\mathbf{p}^{2}}{2m}t+i\frac{\mathbf{px}}{\hbar }\right) d%
\mathbf{p}  \label{h0.2}
\end{equation}
where $\left| \psi _{\mathbf{p}}\right| $ and $\zeta \left( \mathbf{p}%
\right) $ are arbitrary real functions of $\mathbf{p}$. The quantity $\left|
\psi _{\mathbf{p}}\right| \exp \left( i\frac{\mathbf{p}^{2}}{2m}t\right)
\exp \left( -i\zeta \left( \mathbf{p}\right) \right) $ is the Fourier
component of the wave function $\psi \left( t,\mathbf{x}\right) $. The
module $\left| \psi _{\mathbf{p}}\right| $ of the Fourier component of the
wave function is conserved. It can be determined by means of the diffraction
grating. We obtain the momentum distribution $w\left( \mathbf{p}\right) $ by
the relation 
\begin{equation}
w\left( \mathbf{p}\right) =A\left| \psi _{\mathbf{p}}\right| ^{2},\qquad
A=\left( \int \left| \psi _{\mathbf{p}}\right| ^{2}d\mathbf{p}\right) ^{-1}
\label{h0.3}
\end{equation}
The diffraction experiment needs a long time. During this time the phase $i%
\frac{\mathbf{p}^{2}}{2m}t-i\zeta \left( \mathbf{p}\right) $ of the wave
function changes essentially, and one cannot determine to what time and to
what wave function the measured distribution (\ref{h0.3}) should be
attributed. Other methods of the momentum measurement need also a long time 
\cite{R77}.

In fact, it means that the particle momentum is not always described by the
operator $-i\hbar \mathbf{\nabla }$. Operator $-i\hbar \mathbf{\nabla }$ is
the momentum operator only in the case, when the wave function has the form
of the wave packet (\ref{h0.2}), whose space width is large enough, and $%
\left| \psi _{\mathbf{p}}\right| $ does not vanish only in a very small
domain of values of $\mathbf{p}$. In other cases the question whether the
operator $-i\hbar \mathbf{\nabla }$ is the particle momentum operator is
open.

In the present paper we show two points:

\begin{enumerate}
\item  The quantum mechanical technique describes dynamics of statistical
ensembles

\item  The distribution (\ref{h0.2}) disagrees with the von Neumann's
postulate (\ref{h0.1}).
\end{enumerate}

\section{Whose state does the wave function describe?}

The fact, that the Schr\"{o}dinger equation for the free quantum particle
may be written in the form of the hydrodynamic equations, describing an
irrotational flow of some fluid, was known long ago \cite{M26,B26}.
Representation of the Schr\"{o}dinger equation in the form of hydrodynamic
equation for the ideal (quantum) fluid with the internal energy 
\begin{equation}
E=\frac{\hbar ^{2}}{8m\rho }\left( \mathbf{\nabla }\rho \right) ^{2}
\label{h0.3a}
\end{equation}
known as the hydrodynamic interpretation of quantum mechanics, was used by
many authors \cite{M26,B26,B52,T52,T53,JZ63,JZ64,HZ69,B73,BH89,H93}. Here $%
\rho $ is the fluid density, $m$ is the mass of the quantum particle, $%
E=E\left( \rho ,\mathbf{\nabla }\rho \right) $ is the internal energy of the
fluid per unit mass, and $\hbar $ is the quantum constant.

Hydrodynamic description was used mainly as an interpretation of quantum
mechanics, but not as a method of the quantum particle description, because
the hydrodynamic equations are nonlinear and more difficult for solution,
than the Schr\"{o}dinger equation. On the other hand, the hydrodynamic
interpretation is more demonstrative, than the conventional interpretation
in terms of the wave function.

Although the connection between the Schr\"{o}dinger equation and the
hydrodynamic description was known for many years, all this time it was the
one-way connection. One could obtain the irrotational fluid flow from the
dynamic equation for the wave function (Schr\"{o}dinger equation), but one
did not know how to transform dynamic equations for a fluid to the dynamic
equation for a wave function. In other words, we did not know how to
describe rotational fluid flow in terms of the wave function. In terms of
the wave function we could describe only irrotational fluid flow.

To describe an arbitrary fluid flow in terms of a wave function, one needs 
\textit{to integrate conventional dynamic equations for a fluid} (Euler
equations). Indeed, the Schr\"{o}dinger equation 
\begin{equation}
i\hbar \frac{\partial \psi }{\partial t}+\frac{\hbar ^{2}}{2m}\mathbf{\nabla 
}^{2}\psi =0  \label{a1.6}
\end{equation}
may be reduced to the hydrodynamic equations for the density $\rho $ and the
velocity $\mathbf{v}$ of some ideal (quantum) fluid. Substituting $\psi =%
\sqrt{\rho }\exp \left( i\hbar \varphi \right) $ in (\ref{a1.6}) and
separating real and imaginary parts of the equation, we obtain expressions
for the time derivatives $\partial _{0}\rho $ and $\partial _{0}\varphi $.
To obtain expression for the time derivative $\partial _{0}\mathbf{v}$ of
the velocity $\mathbf{v=}\frac{\hbar }{m}\mathbf{\nabla }\varphi $, we need
to differentiate the dynamic equation for $\partial _{0}\varphi $, forming
the combination $\partial _{0}\mathbf{v=\nabla }\left( \frac{\hbar }{m}%
\partial _{0}\varphi \right) $. The reverse transition from the hydrodynamic
equations to the dynamic equations for the wave function (Schr\"{o}dinger
equation) needs a general integration of hydrodynamic equations. This
integration is simple in the partial case of the irrotational flow, but it
is a rather complicated mathematical problem in the general case, when the
result of integration has to contain three arbitrary functions of three
arguments. Without producing this integration, one cannot derive a
description of a fluid in terms of the wave function, and one cannot
manipulate the dynamic equations, transforming them from the representation
in terms of $\rho $, $\mathbf{v}$ to the representation in terms of the wave
function and back. This problem has not been solved for years. Now it has
been solved. It has been shown that any ideal fluid can be described in
terms of a many-component complex wave function. The irrotational flow of
the ideal fluid is described by the one-component wave function, whereas the
rotational flow is described by the two-component wave function (or even by
the many-component one) \cite{R99}. It means that the wave function and the
spin are attributes of the ideal fluid description (but not those of the
quantum principles). In other words, the wave function is only a way of
description of the quantum fluid, whereas the properties of the quantum
phenomena are determined by the quantum fluid itself, whose properties are
determined in turn by its internal energy. It means that the quantum fluid
is a real vehicle of quantum properties, but not the wave function, which is
simply a way of the ideal fluid description.

\textit{Any statistical description }contains two different objects:
individual stochastic object $\mathcal{S}_{\mathrm{st}}$ (individual
particle) and statistical average object (particle) $\left\langle \mathcal{S}%
_{\mathrm{st}}\right\rangle $. The individual stochastic particle is a
stochastic system $\mathcal{S}_{\mathrm{st}}$. The state of the system $%
\mathcal{S}_{\mathrm{st}}$ is described by its position $\mathbf{x}$ and its
velocity $\mathbf{v}$. By definition, the stochasticity of $\mathcal{S}_{%
\mathrm{st}}$ means that a single measurement ($S$-measurement) of a state
of $\mathcal{S}_{\mathrm{st}}$ is irreproducible, i.e. preparing the
particle $\mathcal{S}_{\mathrm{st}}$ in the same way and repeating the $S$%
-measurement (of the particle position), we obtain, in general, another
result, because there are no dynamic equations for $\mathcal{S}_{\mathrm{st}%
} $. But the evolution of the individual system $\mathcal{S}_{\mathrm{st}}$
is not stochastic completely. It contains some regular evolution component $%
\mathcal{C}_{\mathrm{reg}}$.

To obtain $\mathcal{C}_{\mathrm{reg}}$, we consider a set $\mathcal{E}\left[
N,\mathcal{S}_{\mathrm{st}}\right] $ of $N$ independent identical stochastic
systems $\mathcal{S}_{\mathrm{st}}$. The set $\mathcal{E}\left[ N,\mathcal{S}%
_{\mathrm{st}}\right] $, known as the statistical ensemble, is a stochastic
system, having $6N$ degrees of freedom (it is supposed that $\mathcal{S}_{%
\mathrm{st}}$ has $6$ degrees of freedom). If $N$ is large enough, the
stochastic components $\mathcal{C}_{\mathrm{st}}$ of the state evolution
compensate each other, whereas the regular components $\mathcal{C}_{\mathrm{%
reg}}$ are accumulated, and the statistical ensemble $\mathcal{E}\left[ N,%
\mathcal{S}_{\mathrm{st}}\right] $, ($N\rightarrow \infty $) becomes to be a
dynamic system. In the limit $N\rightarrow \infty $ the set $\mathcal{E}%
\left[ N,\mathcal{S}_{\mathrm{st}}\right] $ turns to the statistical
ensemble $\mathcal{E}\left[ \infty ,\mathcal{S}_{\mathrm{st}}\right] $,
which is a continuous dynamic system, having infinite number of the freedom
degrees. There are dynamic equations for $\mathcal{E}\left[ \infty ,\mathcal{%
S}_{\mathrm{st}}\right] $, which can be obtained as a result of variation of
the action functional $\mathcal{A}_{\mathcal{E}\left[ \infty ,\mathcal{S}_{%
\mathrm{st}}\right] }$. All essential characteristics of the statistical
ensemble $\mathcal{E}\left[ N,\mathcal{S}_{\mathrm{st}}\right] $ do not
depend on the number $N$ of elements $\mathcal{S}_{\mathrm{st}}$ of the
statistical ensemble $\mathcal{E}\left[ N,\mathcal{S}_{\mathrm{st}}\right] $%
, if $N$ is large enough, and one can normalize the statistical ensemble to
one system and introduce the statistical average system $\left\langle 
\mathcal{S}_{\mathrm{st}}\right\rangle $, which is the dynamical system,
whose action functional $\mathcal{A}_{\left\langle \mathcal{S}_{\mathrm{st}%
}\right\rangle }$ is defined by the relation 
\begin{equation}
\mathcal{A}_{\left\langle \mathcal{S}_{\mathrm{st}}\right\rangle
}=\lim_{N\rightarrow \infty }\frac{1}{N}\mathcal{A}_{\mathcal{E}\left[ N,%
\mathcal{S}_{\mathrm{st}}\right] }  \label{h1.1}
\end{equation}
Thus, the statistical average system $\left\langle \mathcal{S}_{\mathrm{st}%
}\right\rangle $ is a continuous dynamic system with the infinite number of
the freedom degrees. Investigating $\left\langle \mathcal{S}_{\mathrm{st}%
}\right\rangle $, we study the regular evolution component $\mathcal{C}_{%
\mathrm{reg}}$ of the stochastic system $\mathcal{S}_{\mathrm{st}}$, and the
hydrodynamic description of $\left\langle \mathcal{S}_{\mathrm{st}%
}\right\rangle $ is a way of the investigation of $\mathcal{S}_{\mathrm{st}}$%
.

Although ACQP is a statistical conception, it contains only one sort of
objects: quantum particle (system) $\mathcal{S}_{\mathrm{q}}$. The quantum
system $\mathcal{S}_{\mathrm{q}}$ is a continuous dynamic system, whose
state (wave function $\psi $) is a point in the infinite-dimensional Hilbert
space. ACQP pretends to be a special kind of statistical conception, which
contains only one sort of physical objects: quantum particle $\mathcal{S}_{%
\mathrm{q}}$. According to conventional interpretation of the quantum
mechanics \cite{N32} the quantum particle $\mathcal{S}_{\mathrm{q}}$ is an
individual particle $\mathcal{S}_{\mathrm{st}}$, but not the statistical
average particle $\left\langle \mathcal{S}_{\mathrm{st}}\right\rangle $.
Such an identification generates the puzzling question: '' The state of the
deterministic classical particle $\mathcal{S}_{\mathrm{d}}$ is a point in
six-dimensional phase space. The state of the statistical ensemble $\mathcal{%
E}\left[ N,\mathcal{S}_{\mathrm{d}}\right] $ of $N$ ($N\rightarrow \infty )$
deterministic particles $\mathcal{S}_{\mathrm{d}}$ is described as a point
in $6N$-dimensional phase space. Why is the state of the individual quantum
system $\mathcal{S}_{\mathrm{q}}$ described as a point in the
infinite-dimensional Hilbert space?''

From the viewpoint of the quantum mechanical technique it would be better to
identify the individual quantum particle $\mathcal{S}_{q}$ with the
statistical average particle $\left\langle \mathcal{S}_{\mathrm{st}%
}\right\rangle $, whose state is also a point in the infinite-dimensional
space. But the answer of ACQP is as follows: ''The quantum mechanics is a
special kind of the statistical conception, which contains only one sort of
objects $\mathcal{S}_{\mathrm{q}}$, but not a partial case of the general
statistical conception, containing two sorts of objects $\mathcal{S}_{%
\mathrm{st}}$ and $\left\langle \mathcal{S}_{\mathrm{st}}\right\rangle $.
The state of individual quantum particle $\mathcal{S}_{\mathrm{q}}$ is
described by the wave function $\psi $, and it is a postulate of the quantum
mechanics.''

Another question is as follows. Let us set $\hbar =0$ in the description of
the quantum particle $\mathcal{S}_{\mathrm{q}}$. We must obtain a
description of the deterministic classical particle. What is this
description? A description of individual particle $\mathcal{S}_{\mathrm{d}}$%
, or a description of the statistical ensemble $\mathcal{E}\left[ \infty ,%
\mathcal{S}_{\mathrm{d}}\right] $ of classical particles $\mathcal{S}_{%
\mathrm{d}}$? At first, let us note that one cannot set $\hbar =0$ in the
Schr\"{o}dinger equation (\ref{a1.6}), because in this case we do not obtain
any description. Before setting $\hbar =0$, we are to transform the phase of
the wave function, setting 
\begin{equation}
\psi =\sqrt{\rho }\exp \left( i\varphi \right) =\sqrt{\rho }\exp \left( 
\frac{i}{\hbar }S\right)  \label{h0.4}
\end{equation}
Substituting (\ref{h0.4}) in (\ref{a1.6}) and separating real and imaginary
parts of the equation, we obtain 
\begin{equation}
\frac{i}{2\rho }\left( \frac{\partial \rho }{\partial t}+\frac{\rho }{m}%
\nabla ^{2}S+\frac{\nabla \rho }{m}\nabla S\right) =0  \label{h0.5}
\end{equation}
\begin{equation}
\frac{\partial S}{\partial t}-\frac{\left( \mathbf{\nabla }S\right) ^{2}}{2m}%
+\frac{\hbar ^{2}}{2m}\left( \mathbf{\nabla }\frac{\mathbf{\nabla }\rho }{%
2\rho }+\left( \frac{\mathbf{\nabla }\rho }{2\rho }\right) ^{2}\right) =0
\label{h0.6}
\end{equation}
Now let us introduce designation $\mathbf{v}=m^{-1}\mathbf{\nabla }S$, take
the gradient of the equation (\ref{h0.6}) and set $\hbar =0$ in (\ref{h0.5})
and (\ref{h0.6}). We obtain 
\begin{equation}
\frac{\partial \rho }{\partial t}+\mathbf{\nabla }\left( \rho \mathbf{v}%
\right) =0,\qquad \frac{\partial \mathbf{v}}{\partial t}+\left( \mathbf{%
v\nabla }\right) \mathbf{v}=0  \label{h0.7}
\end{equation}

Thus, we obtain dynamic equations for the ideal fluid without a pressure.
This fluid is the statistical ensemble $\mathcal{E}\left[ \infty ,\mathcal{S}%
_{\mathrm{d}}\right] $ of classical particles $\mathcal{S}_{\mathrm{d}}.$ We
see that at $\hbar =0$ the quantum system $\mathcal{S}_{\mathrm{q}}$ turns
to a statistical ensemble, but not to an individual system. It means that
from the formal viewpoint the quantum particle $\mathcal{S}_{\mathrm{q}}$
should be regarded as the statistical ensemble $\mathcal{E}\left[ \infty ,%
\mathcal{S}_{\mathrm{st}}\right] $, or as the statistical average particle $%
\left\langle \mathcal{S}_{\mathrm{st}}\right\rangle $, but not as an
individual particle.

Another interesting question: ''Why does one need to transform the scale of
the wave function phase $\varphi $, introducing factor $\hbar ^{-1}$ in the
exponent of (\ref{h0.4})?'' At $\hbar \rightarrow 0$ this factor tends to
infinity. The answer is rather curious. The action functional, describing
the irrotational flow of the ideal fluid with the internal energy (\ref
{h0.3a}) has the form \cite{R99,R002} 
\begin{equation}
\mathcal{A}[\psi ,\psi ^{\ast }]=\int \left\{ \frac{ib_{0}}{2}(\psi ^{\ast
}\partial _{0}\psi -\partial _{0}\psi ^{\ast }\cdot \psi )-\frac{b_{0}^{2}}{%
2m}\mathbf{\nabla }\psi ^{\ast }\cdot \mathbf{\nabla }\psi +\frac{b_{0}^{2}}{%
8m}\frac{\left( \mathbf{\nabla }\rho \right) ^{2}}{\rho }-\frac{\hbar ^{2}}{%
8m}\frac{\left( \mathbf{\nabla }\rho \right) ^{2}}{\rho }\right\} \mathrm{d}%
^{4}x,  \label{h0.8}
\end{equation}
where $\rho =\psi ^{\ast }\psi $, $\psi $ is a complex one-component wave
function, $\psi ^{\ast }$ is the quantity complex conjugate to $\psi $. The
quantity $b_{0}$ is an arbitrary (integration) constant ($b_{0}\neq 0$),
describing the scale of the wave function phase. The last term of (\ref{h0.8}%
) describes the internal energy of the fluid. Only this term contains the
quantum constant $\hbar $. All other terms are dynamical terms which are
present at the description of any nonrelativistic statistical ensemble.
Dynamic equation generated by the action (\ref{h0.8}) is equivalent to the
Schr\"{o}dinger equation for any choice of the constant $b_{0}$. If we set $%
\hbar =0$, the last term in (\ref{h0.8}) vanishes, the internal energy of
the fluid becomes $E\left( \rho ,\mathbf{\nabla }\rho \right) =0$. In this
case the action (\ref{h0.8}) describes the statistical ensemble of classical
(deterministic) particles $\mathcal{S}_{\mathrm{d}}$. Thus, in the action (%
\ref{h0.8}) the transition to the classical case is obtained by setting $%
\hbar =0$.

Formal variation of the action (\ref{h0.8}) with respect to $\psi ^{\ast }$
leads to the dynamic equation for $\psi $, which is nonlinear because of two
last terms in (\ref{h0.8}). If we chose the integration constant $%
b_{0}=\hbar $, two last terms compensate each other and the action (\ref
{h0.8}) turns to the action 
\begin{equation}
\mathcal{A}[\psi ,\psi ^{\ast }]=\int \left\{ \frac{i\hbar }{2}(\psi ^{\ast
}\partial _{0}\psi -\partial _{0}\psi ^{\ast }\cdot \psi )-\frac{\hbar ^{2}}{%
2m}\mathbf{\nabla }\psi ^{\ast }\cdot \mathbf{\nabla }\psi \right\} \mathrm{d%
}^{4}x  \label{h0.9}
\end{equation}
which generates the Schr\"{o}dinger equation (\ref{a1.6}) directly. But now
all terms of the action (\ref{h0.9}) are quantum in the sense that they
contain the quantum constant $\hbar $ as a factor. Setting $\hbar =0$, we
cannot transit now to the classical case, because at $\hbar =0$ the action (%
\ref{h0.9}) vanishes. However, we can transform the phase scale of $\psi $
by means of the transformation \cite{R002}

\begin{equation}
\psi \rightarrow \tilde{\psi}=|\psi |\exp \left( \frac{b_{0}}{\hbar }\log 
\frac{\psi }{|\psi |}\right) ,  \label{h0.10}
\end{equation}
After this transformation the action (\ref{h0.9}) turns to the action (\ref
{h0.8}). If we now set $\hbar =0$, we obtain the action for the statistical
ensemble of classical particles. Making the change of the phase $\varphi $ (%
\ref{h0.4}) in the Schr\"{o}dinger equation (\ref{a1.6}), we produce the
transformation (\ref{h0.10}) with $b_{0}=1$. Thereafter setting $\hbar =0$,
we can obtain dynamic equations for the statistical ensemble of classical
particles.

Thus, the formal consideration shows that the quantum particle $\mathcal{S}_{%
\mathrm{q}}$ is in reality the statistical average particle $\left\langle 
\mathcal{S}_{\mathrm{st}}\right\rangle $, or the statistical ensemble $%
\mathcal{E}\left[ \mathcal{S}_{\mathrm{st}}\right] $, but not an individual
particle. If we consider $\mathcal{S}_{\mathrm{q}}$ as an individual
particle, we destroy the continuous connection between the classical physics
and the quantum mechanics.

Some authors consider the quantum particle $\mathcal{S}_{\mathrm{q}}$ as the
statistical ensemble $\mathcal{E}\left[ \mathcal{S}_{\mathrm{st}}\right] $
of stochastic particles $\mathcal{S}_{\mathrm{st}}$ (for instance \cite{B76}%
). Such an interpretation is better, because it reestablishes the continuous
connection between the classical physics and the quantum mechanics.
Unfortunately, the quantum mechanics is presented not always consistently,
because only one sort of objects (particle $\mathcal{S}_{\mathrm{q}}=%
\mathcal{E}\left[ \mathcal{S}_{\mathrm{st}}\right] $) is considered. Such an
interpretation is in accordance with the quantum mechanical technique, which
deals with only one sort of particles $\mathcal{S}_{\mathrm{q}}=\mathcal{E}%
\left[ \mathcal{S}_{\mathrm{st}}\right] $, and all quantum mechanical
predictions concern only with $\mathcal{S}_{\mathrm{q}}=\mathcal{E}\left[ 
\mathcal{S}_{\mathrm{st}}\right] $. But there are two sorts of measurements:
(1) the single measurement ($S$-measurement) produced under an individual
particle and (2) the massive measurement ($M$-measurement) produced under a
statistical ensemble. Properties of $S$-measurement and those of $M$%
-measurement are discrepant and incompatible. The result of the $S$%
-measurement is a definite value of the measured quantity, whereas the
result of the $M$-measurement is a distribution of the measured quantity.
The result of the $S$-measurement is random and irreproducible, whereas the
result of the $M$-measurement is regular and reproducible. The $S$%
-measurement does not change the state (wave function), whereas the $M$%
-measurement replaces the wave function $\psi $ by the density matrix. If we
do not distinguish between the $M$-measurement and $S$-measurement and use
one term ''measurement'' for the two kinds of measurement, it is a source of
possible paradoxes \cite{R002c}.

Unfortunately, we have not met such a presentation of the quantum mechanics,
where two kinds of measurements be considered. Such a distinction between
two kinds of measurements is necessary, if we consider the quantum mechanics
as a special case of the general statistical conception (dynamics of
stochastic systems). We have not met such a distinction between $S$%
-measurement and $M$-measurement even in the presentation of those authors,
who believe that the wave function describes the state of the statistical
ensemble (but not of a single particle). The presentation of quantum
mechanics cannot be consistent without this distinction. Besides, any $M$%
-measurement is a set of many $S$-measurements, and one may not ignore $S$%
-measurement, although its results are not described by the QM formalism.

The quantum mechanics is a logical construction which contains axioms
(primary statements) and corollaries of these axioms. The primary statements
can be chosen in different way, and interpretation of quantum mechanics
depends on this choice. This choice is unessential for the nonrelativistic
quantum mechanics in itself, but it is essential, if we are going to expand
the quantum mechanics onto the relativistic quantum phenomena. For instance,
if the quantum principles are primary statements of the quantum mechanics,
then dynamic equations in terms of the wave function must be linear, and the
internal energy of the quantum fluid must have the form (\ref{h0.3a}), in
order the last term of (\ref{h0.8}) may be compensated with the antecedent
term at $\hbar =b_{0}$. Any other form of the internal energy fluid is
unallowable. Compensation of the dynamical term with the term describing
some special fluid property looks rather artificial. It is justified only by
the intention to obtain the linear dynamic equation. From the point of view
of the stochastic system dynamics, where all ideal fluids are admissible,
the consideration of the dynamic equation linearity as a principle of the
logical construction seems to be rather doubtful. Of course, if for some
special quantum fluid we can obtain the linear dynamic equation, be it even
an artificial identification of the terms of different nature, we should
accept and use this identification. But it seems doubtful to consider the
linearity connected with such an identification as a principle of the
quantum mechanics.

Besides, the action for the quantum fluid with internal energy (\ref{h0.3a})
has the form (\ref{h0.8}) only for irrotational flow of the fluid. The wave
function $\psi $ is two-component at the rotational flow of the same fluid,
and the action has the form \cite{R99} 
\begin{eqnarray}
\mathcal{A}[\psi ,\psi ^{\ast }] &=&\int \left\{ \frac{ib_{0}}{2}(\psi
^{\ast }\partial _{0}\psi -\partial _{0}\psi ^{\ast }\cdot \psi )-\frac{%
b_{0}^{2}}{2m}\mathbf{\nabla }\psi ^{\ast }\cdot \mathbf{\nabla }\psi \right.
\nonumber \\
&&+\left. \frac{b_{0}^{2}}{8m}(\mathbf{\nabla }s_{\alpha })(\mathbf{\nabla }%
s_{\alpha })\rho +\frac{b_{0}^{2}-\hbar ^{2}}{8m}\frac{\left( \mathbf{\nabla 
}\rho \right) ^{2}}{\rho }\right\} \mathrm{d}^{4}x  \label{h2.12a}
\end{eqnarray}
\begin{equation}
\psi =\left( _{\psi _{2}}^{\psi _{1}}\right) ,\qquad \psi ^{\ast }=\left(
\psi _{1}^{\ast },\psi _{2}^{\ast }\right) ,\qquad \rho \equiv \psi ^{\ast
}\psi ,\qquad s_{\alpha }\equiv \frac{{\psi ^{\ast }}\sigma _{\alpha }{\psi }%
}{\rho },\qquad \alpha =1,2,3,  \label{h0.12}
\end{equation}
where $\sigma _{\alpha }$ are the Pauli matrices. Dynamic equation for the
rotational fluid flow is nonlinear, even if $b_{0}=\hbar $. It has the form 
\begin{equation}
i\hbar \partial _{0}\psi +\frac{\hbar ^{2}}{2m}\partial _{\alpha }\partial
_{\alpha }\psi +\frac{\hbar ^{2}}{8m}\left( \mathbf{\nabla }s_{\alpha
}\right) \left( \mathbf{\nabla }s_{\alpha }\right) \psi =\hbar ^{2}\frac{%
\mathbf{\nabla }\left( \rho \mathbf{\nabla }s_{\alpha }\right) }{4\rho m}%
\left( \sigma _{\alpha }-s_{\alpha }\right) \psi  \label{h0.14}
\end{equation}
This equation is nonlinear, and one can hardly reduce it to a linear
equation. Then the equation (\ref{h0.14}) is incompatible with principles of
the quantum mechanics.

Being considered as a principle, the linearity of the dynamic equation
restricts strongly capacity of the stochastic system dynamics. In
particular, attempts of expansion of the quantum mechanics on relativistic
phenomena of microcosm failed, and the constraints imposed on the dynamics
of the stochastic systems by the quantum mechanics principles may play an
essential role in this failure.

In the conventional (Copenhagen) interpretation of quantum mechanics the
quantum particle motion is described in terms of the wave function $\psi $
and related concepts: interference, diffraction, coherence. Such concepts as
momentum, energy, angular momentum are used also, but they are expressed via
the wave function $\psi $, and their sense is sometimes another, than in the
classical mechanics. But the wave function $\psi $ is a nonrelativistic
concept, and the conventional interpretation meets difficulties at its
expansion onto the relativistic motion of quantum particles. It is the main
defect of the conventional interpretation. Besides, the wave function is
only a way of the quantum fluid description. It is unreliable to construct
the interpretation on the basis of some method of a description.
Consideration of the quantum fluid (that is $\left\langle \mathcal{S}_{%
\mathrm{st}}\right\rangle $) in itself is a more reliable interpretation.
Being a fluidlike continuous dynamic system, the statistical average
particle $\left\langle \mathcal{S}_{\mathrm{st}}\right\rangle $ can be
described in terms of usual hydrodynamic concepts: the flux 4-vector $%
j^{i}=\left\{ \rho ,\rho \mathbf{v}\right\} $ and the energy-momentum tensor 
$T^{ik}$.

Integrating the system of dynamic equations

\begin{equation}
\frac{dx^{i}}{d\tau }=j^{i}\left( x\right) ,\qquad i=0,1,2,3  \label{h1.2}
\end{equation}
where 
\begin{equation}
j^{i}=\left\{ \rho ,\mathbf{j}\right\} ,\qquad j^{0}=\rho =\psi ^{\ast }\psi
,\qquad \mathbf{j}=\frac{\hbar }{2m}\left( \psi ^{\ast }\mathbf{\nabla }\psi
-\mathbf{\nabla }\psi ^{\ast }\cdot \psi \right)  \label{h.2a}
\end{equation}
we obtain the mean world lines $\mathcal{L}:x^{i}=x^{i}\left( \tau \right) $%
, associated with the statistical average particle $\left\langle \mathcal{S}%
_{\mathrm{st}}\right\rangle $. Considering the energy-momentum tensor $%
T^{ik} $ along $\mathcal{L}$, we can evaluate the energy-momentum
characteristics associated with the mean world line $\mathcal{L}$. Thus, the
hydrodynamic interpretation describes all quantum phenomena in terms of the
mean world lines and other attributes of hydrodynamics.

Properties of the quantum fluid are determined by the internal energy $E$ of
the fluid, which depends on the properties of some 4-vector field $u^{k}$, $%
k=0,1,2,3$. This vector describes the mean value of the stochastic component
of the particle velocity. Choice of the field of this vector $u^{k}$ means
the procedure of the classical particle quantization. To obtain the action $%
\mathcal{A}_{\left\langle \mathcal{S}_{\mathrm{st}}\right\rangle }$ for the
quantum fluid $\left\langle \mathcal{S}_{\mathrm{st}}\right\rangle $, it is
necessary to take the action $\mathcal{A}_{\mathcal{E}_{\mathrm{d}}\left[ 
\mathcal{S}_{\mathrm{d}}\left( P\right) \right] }$ for the statistical
ensemble $\mathcal{E}_{\mathrm{d}}\left[ \mathcal{S}_{\mathrm{d}}\left(
P\right) \right] $ of deterministic classical particles $\mathcal{S}_{%
\mathrm{d}}\left( P\right) $, where $P$ is a set of parameters of the
classical particle. For the free classical particle there is only one
parameter $P$: its mass $m$. The action for the ensemble $\mathcal{E}_{%
\mathrm{d}}\left[ \mathcal{S}_{\mathrm{d}}\right] $ of free deterministic
particles $\mathcal{S}_{\mathrm{d}}$ has the form 
\begin{equation}
\mathcal{E}_{\mathrm{d}}\left[ \mathcal{S}_{\mathrm{d}}\right] :\qquad 
\mathcal{A}_{\mathcal{E}_{\mathrm{d}}\left[ \mathcal{S}_{\mathrm{d}}\left(
m\right) \right] }\left[ \mathbf{x}\right] =\int L\left( \mathbf{x,}\frac{d%
\mathbf{x}}{dt}\right) dtd\mathbf{\xi },  \label{a0.9}
\end{equation}
where the Lagrangian function density is described by the relation 
\begin{equation}
L\left( \mathbf{x,}\frac{d\mathbf{x}}{dt}\right) =-mc^{2}+\frac{m}{2}\left( 
\frac{d\mathbf{x}}{dt}\right) ^{2}  \label{a0.8}
\end{equation}
$\mathbf{x}=\mathbf{x}\left( t,\mathbf{\xi }\right) $, $\mathbf{\xi }%
=\left\{ \xi _{1},\xi _{2},\xi _{3}\right\} $. Here variables $\mathbf{\xi }$
label the particles $\mathcal{S}_{\mathrm{d}}$ of the statistical ensemble $%
\mathcal{E}_{\mathrm{d}}\left[ \mathcal{S}_{\mathrm{d}}\right] $. The
constant $c$ is the speed of the light. The action (\ref{a0.9}) describes
some fluid without pressure. Let us replace now the value of $m$ in (\ref
{a0.8}), (\ref{a0.9}) by its effective value $m_{\mathrm{eff}}$ 
\begin{equation}
m\rightarrow m_{\mathrm{eff}}=m\left( 1-\frac{\mathbf{u}^{2}}{2c^{2}}+\frac{%
\hbar }{2mc^{2}}\mathbf{\nabla u}\right)  \label{a0.9a}
\end{equation}
The change (\ref{a0.9a}) is nonrelativistic. In this case the component $%
u^{0}$ of the 4-vector $u^{k}$ is small and it can be neglected \cite{R002}.
In the nonrelativistic approximation we obtain the action $\mathcal{E}_{%
\mathrm{st}}\left[ \mathcal{S}_{\mathrm{st}}\right] $ for the statistical
ensemble of stochastic particles $\mathcal{S}_{\mathrm{st}}$ 
\begin{equation}
\mathcal{E}_{\mathrm{st}}\left[ \mathcal{S}_{\mathrm{st}}\right] :\qquad 
\mathcal{A}_{\mathcal{E}_{\mathrm{d}}\left[ \mathcal{S}_{\mathrm{d}}\left(
m_{\mathrm{eff}}\right) \right] }\left[ \mathbf{x,u}\right] =\int L\left( 
\mathbf{x,}\frac{d\mathbf{x}}{dt}\right) +L_{\mathrm{st}}\left( \mathbf{u},%
\mathbf{\nabla u}\right) dtd\mathbf{\xi },  \label{a0.10}
\end{equation}
\begin{equation}
L_{\mathrm{st}}\left( \mathbf{u},\mathbf{\nabla u}\right) =\frac{m}{2}%
\mathbf{u}^{2}-\frac{\hbar }{2}\mathbf{\nabla u}  \label{a0.11}
\end{equation}
Here $\mathbf{u}=\mathbf{u}\left( t,\mathbf{x}\right) $ is the mean value of
the stochastic velocity component. The first term in (\ref{a0.11}) describes
the energy of the stochastic component of the velocity. The quantum constant 
$\hbar $ appears here as a coupling constant, describing connection between
the regular and stochastic components of the particle motion. The velocity $%
\mathbf{u}$ is supposed to be small with respect to $c$. Dynamic equations
for the quantum fluid with the pressure are obtained as a result of
variation of the action (\ref{a0.10}) with respect to variables $\mathbf{x}$
and $\mathbf{u}$. It follows from the dynamic equation for $\mathbf{u}$ that 
\begin{equation}
\mathbf{u=-}\frac{\hbar }{2m}\mathbf{\nabla }\ln \rho ,\qquad E_{\mathrm{int}%
}=\frac{m\mathbf{u}^{2}}{2}\rho =\frac{\hbar ^{2}}{8m}\frac{\left( \mathbf{%
\nabla }\rho \right) ^{2}}{\rho },\qquad \rho =\left( \frac{\partial \left(
x^{1},x^{2},x^{3}\right) }{\partial \left( \xi _{1},\xi _{2},\xi _{3}\right) 
}\right) ^{-1}  \label{h1.6}
\end{equation}
where $\rho $ is the fluid density, and $E_{\mathrm{int}}$ is the energy
density connected with the mean value $\mathbf{u}$ of the stochastic
velocity component. One can show that the irrotational flow of the quantum
fluid is described by the Schr\"{o}dinger equation (\ref{a1.6}) \cite{R002}.
Thus, the quantization of the classical particle is carried out by an
addition of a supplemental term (\ref{a0.11}) to the action $\mathcal{E}_{%
\mathrm{d}}\left[ \mathcal{S}_{\mathrm{d}}\right] $ for the statistical
ensemble of free classical particles $\mathcal{S}_{\mathrm{d}}$. The
statistical ensemble $\mathcal{E}_{\mathrm{d}}\left[ \mathcal{S}_{\mathrm{d}}%
\right] $ ceases to be a statistical ensemble of classical particles $%
\mathcal{S}_{\mathrm{d}}$, because now its elements interact between
themselves. This interaction generates a pressure in the quantum fluid.

Quantization of free relativistic classical particles is carried out by
means of the relativistic version of the change (\ref{a0.9a}) \cite{R99,R98}%
. Sometimes a puzzling question arises: ''How can the quantum fluid describe
such wave phenomena as interference and diffraction?'' The fact is that the
pressure and the internal energy of the quantum fluid depend on the density $%
\rho $ and on the density gradient $\mathbf{\nabla }\rho $, whereas in the
usual fluid they depend only on the density $\rho $. Appearance of
additional spatial derivatives in the hydrodynamic equations generates the
wave properties of the quantum fluid. Formally it follows from the
description of the irrotational flow of quantum fluid in terms of the
Schr\"{o}dinger equation.

\section{The momentum distribution}

Hydrodynamic interpretation of quantum mechanics cannot give distribution $%
w\left( \mathbf{p}\right) $ over momenta $\mathbf{p}$ at the state,
described by the wave function $\psi $. It can give only the mean momentum
distribution. The conventional (Copenhagen) interpretation gives the
momentum distribution by the relation

\begin{equation}
w\left( \mathbf{p}\right) =A\psi _{\mathbf{p}}^{\ast }\psi _{\mathbf{p}%
},\qquad \psi _{\mathbf{p}}=\frac{1}{\left( 2\pi \hbar \right) ^{3}}\int
\psi \left( \mathbf{x}\right) e^{-i\frac{\mathbf{px}}{\hbar }}d\mathbf{%
x\qquad A=}\int \psi \mathbf{_{\mathbf{p}}^{\ast }}\psi \mathbf{_{\mathbf{p}}%
}d\mathbf{p,}  \label{a1.1}
\end{equation}
where $\psi _{\mathbf{p}}$ is the Fourier component of the wave function $%
\psi $. But this statement cannot be tested experimentally, because the
measured distribution cannot be attributed to the state $\psi \left( \mathbf{%
x}\right) $. On the contrary, the mean momentum distribution, which is given
by the hydrodynamic interpretation, can be tested experimentally and
attributed to the wave function $\psi $, but this distribution does not
coincide with (\ref{a1.1}).

To realize the difference between the momentum and the mean momentum, we
consider the measurement of the molecule momentum in the example of the
stationary flow of an ideal gas. Let the gas flow be isothermal, and the
temperature $kT=$const. Distribution of the gas molecules over velocities is
described by the Maxwell distribution 
\begin{equation}
F\left( \mathbf{x},\mathbf{v}\right) d\mathbf{v}=\left( \frac{m}{2\pi kT}%
\right) ^{3/2}\exp \left\{ -\frac{m\left( \mathbf{v}-\mathbf{u}\left( 
\mathbf{x}\right) \right) ^{2}}{2kT}\right\} d\mathbf{v}  \label{a1.2}
\end{equation}
where $\mathbf{u}\left( \mathbf{x}\right) $ is the gas velocity at the point 
$\mathbf{x}$ (the mean velocity of the gas molecules at the point $\mathbf{x}
$). The distribution over momenta $\mathbf{p}=m\mathbf{v}$ has the form 
\begin{equation}
F_{1}\left( \mathbf{x},\mathbf{p}\right) d\mathbf{p}=\frac{1}{\left( 2\pi
mkT\right) ^{3/2}}\exp \left\{ -\frac{\left( \mathbf{p}-m\mathbf{u}\left( 
\mathbf{x}\right) \right) ^{2}}{2mkT}\right\} d\mathbf{p}  \label{a1.3}
\end{equation}
Let us divide the volume $V$ of the gas flow into similar cubic cells $%
V_{1},V_{2},...V_{N}$, $N\gg 1$. Let the following conditions be satisfied 
\begin{equation}
l_{\mathrm{c}}\ll L,\qquad \left| v_{\mathrm{t}}\tau _{c}\right| \ll
L,\qquad \left| \mathbf{u}\left( \mathbf{x}\right) \right| \ll v_{\mathrm{t}%
}=\sqrt{\frac{3kT}{m}}  \label{a1.4}
\end{equation}
where $L$ is the linear size of the cell, $l_{\mathrm{c}}$ is the mean path
between the molecule collisions, $\tau _{\mathrm{c}}$ is the mean time
between the collisions and $v_{\mathrm{t}}$ is the mean thermal velocity of
molecules.

Let us calculate the mean momentum $\left\langle \mathbf{p}_{i}\right\rangle 
$ of the gas molecule in the cell $V_{i}$. We obtain $\left\langle \mathbf{p}%
_{i}\right\rangle =m\mathbf{u}\left( \mathbf{x}\right) $, \ $\mathbf{x}\in
V_{i}$, $i=1,2,...N$. The set of all $\left\langle \mathbf{p}%
_{i}\right\rangle $, $i=1,2,...N$ forms the mean momentum distribution. This
distribution is determined completely by the gas flow, and it has nothing to
do with the Maxwell momentum distribution (\ref{a1.3}), which describes both
the regular and random components of the molecule momenta. Under conditions (%
\ref{a1.4}) the mean momentum distribution is much narrower, than the
Maxwell distribution, because the Maxwell distribution takes into account
the random component of the molecule velocity, and in the given case the
random component is much larger, than the regular one. Let us imagine that
we measure the momentum of the accidentally taken molecule. We measure its
position $\mathbf{x}$ at the time $t$ and its position $\mathbf{x}_{1}$ at
the time $t+t_{\mathrm{m}}$. The molecule momentum in the time interval $%
(t,t+t_{\mathrm{m}})$ is determined by the relation 
\begin{equation}
\mathbf{p}=m\frac{\mathbf{x}_{1}-\mathbf{x}}{t_{\mathrm{m}}}  \label{a1.5}
\end{equation}

Producing many such measurements of the molecule momentum, we obtain the
momentum distribution. What distribution do we obtain? The mean momentum
distribution, or the Maxwell distribution? The result depends on the
measurement time $t_{\mathrm{m}}$. If $t_{\mathrm{m}}\ll \tau _{\mathrm{c}}$%
, we obtain the Maxwell distribution (\ref{a1.3}), averaged over the volume $%
V$, occupied by the gas flow. If $\tau _{\mathrm{c}}$ $\ll t_{\mathrm{m}}\ll
L/\left| \mathbf{u}\left( \mathbf{x}\right) \right| $, we obtain the mean
momentum distribution. If $t_{\mathrm{m}}\approx \tau _{\mathrm{c}}$, we
obtain some intermediate result, which is close to the Maxwell distribution,
because the characteristic velocity of the Maxwell distribution $v_{\mathrm{t%
}}=\sqrt{\frac{3kT}{m}}\gg \left| \mathbf{u}\left( \mathbf{x}\right) \right| 
$. Finally, if $\tau _{\mathrm{c}}\ll L/\left| \mathbf{u}\left( \mathbf{x}%
\right) \right| \ll t_{\mathrm{m}}$, we obtain the mean momentum
distribution, containing some additional averaging over the spatial cells.

If the gas flow is nonstationary, the measurement time $t_{\mathrm{m}}$ must
be shorter, than the characteristic time $\tau $ of the gas flow change. If
the mean gas velocity of the nonstationary flow is $\mathbf{u}\left( t,%
\mathbf{x}\right) $, the distribution over the mean momenta $\left\langle 
\mathbf{p}\right\rangle $ of molecules is given by the relation 
\begin{eqnarray}
w\left( \left\langle \mathbf{p}\right\rangle \right) &=&A\int \delta \left(
\left\langle \mathbf{p}\right\rangle -m\mathbf{u}\left( t,\mathbf{x}\right)
\right) \rho \left( t,\mathbf{x}\right) d\mathbf{x,}  \label{h1.3} \\
A &=&\left( \int w\left( \left\langle \mathbf{p}\right\rangle \right)
d\left\langle \mathbf{p}\right\rangle \right) ^{-1}\mathbf{=}\left( \int
\rho \left( t,\mathbf{x}\right) d\mathbf{x}\right) ^{-1}  \nonumber
\end{eqnarray}
where $\rho \left( t,\mathbf{x}\right) $ is the gas density. The mean
momentum $\left\langle \mathbf{p}\right\rangle $ is a single-valued function
of the position $\mathbf{x}$. It is useless to speak on simultaneous
distribution $F\left( \mathbf{x,}\left\langle \mathbf{p}\right\rangle
\right) $ over position and mean momenta, because variables $\mathbf{x}$ and%
\textbf{\ }$\left\langle \mathbf{p}\right\rangle $ are not independent, and
this distribution is a function $\Phi \left( \mathbf{x}\right) $ of only
position $\mathbf{x}$.

Is it possible to obtain distribution over the mean momenta $\left\langle 
\mathbf{p}\right\rangle $ of molecules and to associate it with the state of
the nonstationary gas flow in the case, when $\tau \approx \tau _{\mathrm{c}%
} $? Yes, it is possible. We must measure two subsequent positions $\mathbf{x%
}$ and $\mathbf{x}_{1}$ of a molecule at the times $t$ and $t+t_{\mathrm{m}}$
respectively ($t_{\mathrm{m}}\ll \tau _{\mathrm{c}}\approx \tau $, $\mathbf{x%
}\in V_{i}$) and determine the momentum $\mathbf{p}$ by means of the
relation (\ref{a1.5}). Producing many such measurements at the time $t$ and
taking the mean value of the measured momenta $\mathbf{p}$\textbf{, }we
obtain the mean momentum $\left\langle \mathbf{p}_{i}\right\rangle $ in the
cell $V_{i}$. Producing such measurements at all cells at the moment $t$, we
obtain the distribution over the mean momenta $\left\langle \mathbf{p}%
\right\rangle $ at the time $t$. This mean momentum distribution can be
associated with the state $\rho ,\mathbf{u}$ of the gas flow at the time $t$%
, because the measurement time $t_{\mathrm{m}}\ll \tau \approx \tau _{%
\mathrm{c}}$. On the other hand, this distribution has nothing to do with
the Maxwell distribution over the molecular momenta. Thus, the mean momentum
distribution (\ref{h1.3}) can be measured and attributed to the gas flow
state at the time $t$.

Let us now apply the above consideration to the quantum fluid, describing
the free quantum particle. Let the fluid state be described by the wave
function $\psi $. According to (\ref{h1.3}) we obtain for the mean momentum
distribution 
\begin{equation}
w\left( \left\langle \mathbf{p}\right\rangle \right) =A\int \delta \left(
\left\langle \mathbf{p}\right\rangle -m\mathbf{v}\left( t,\mathbf{x}\right)
\right) \rho \left( t,\mathbf{x}\right) d\mathbf{x,\qquad v}=\frac{\hbar }{%
2m\rho }\left( \psi ^{\ast }\mathbf{\nabla }\psi -\mathbf{\nabla }\psi
^{\ast }\cdot \psi \right) ,  \label{h1.4}
\end{equation}
\begin{equation}
\rho =\psi ^{\ast }\psi ,\qquad \mathbf{A=}\left( \int w\left( \left\langle 
\mathbf{p}\right\rangle \right) d\left\langle \mathbf{p}\right\rangle
\right) ^{-1}\mathbf{=}\left( \int \psi ^{\ast }\psi d\mathbf{x}\right) ^{-1}
\label{h1.5}
\end{equation}
After integration over $\mathbf{x}$ the distribution (\ref{h1.4}) can be
written in the form 
\begin{equation}
w\left( \mathbf{p}\right) =A\sum\limits_{l}\left[ \frac{\rho \left( t,%
\mathbf{x}\right) }{m^{3}\left| D\left( \mathbf{x}\right) \right| }\right] _{%
\mathbf{x}=\mathbf{x}_{l}\left( t,\mathbf{p}\right) },\qquad D\left( \mathbf{%
x}\right) =\frac{\partial \left( v^{1}\left( \mathbf{x}\right) ,v^{2}\left( 
\mathbf{x}\right) ,v^{3}\left( \mathbf{x}\right) \right) }{\partial \left(
x^{1},x^{2},x^{3}\right) }  \label{h1.5a}
\end{equation}
where summation is produced over all roots $\mathbf{x}_{l}\left( t,\mathbf{p}%
\right) $ of the equation 
\begin{equation}
\mathbf{v}\left( t,\mathbf{x}\right) =\frac{\mathbf{p}}{m}  \label{h1.5b}
\end{equation}
The distribution (\ref{h1.4}) can be measured in a very short measurement
time $t_{\mathrm{m}}$ and attributed to some state $\psi $ of the fluid, but
this distribution has not the form (\ref{h0.1}), because it is not bilinear
with respect to the wave function $\psi $. The distribution (\ref{a1.1})
also has not the form (\ref{h0.1}), because it cannot be attributed to the
wave function $\psi $.

In general, the distribution (\ref{h1.4}) distinguishes from the
conventional momentum distribution (\ref{a1.1}). Let us compare the two
distributions for the state, described by the wave function 
\begin{equation}
\psi \left( x\right) =A_{1}e^{-\frac{1}{2}\frac{\left( x-X\right) ^{2}}{a^{2}%
}+\frac{i}{\hbar }k\left( x-X\right) }=\frac{A_{1}\left| a\right| }{\sqrt{%
2\pi }\hbar }\int\limits_{-\infty }^{\infty }e^{-\frac{a^{2}\left(
k-p\right) ^{2}}{2\hbar ^{2}}}e^{\frac{i}{\hbar }p\left( x-X\right)
}dp,\qquad A_{1}=\text{const}  \label{h2.1}
\end{equation}
This wave function describes one-dimensional wave packet of the
characteristic width $a$, moving with the momentum $k$. The center of the
wave packet is placed at the point $X$.

Calculation gives for the distribution (\ref{a1.1}) 
\begin{equation}
w\left( p\right) =A\psi _{p}^{\ast }\psi _{p}=\frac{a}{\sqrt{\pi }\hbar }e^{-%
\frac{a^{2}}{^{\hbar ^{2}}}\left( k-p\right) ^{2}}  \label{h2.2}
\end{equation}

For $\rho $ and $j$ we obtain 
\begin{equation}
\rho \left( x\right) =\left| \psi \right| ^{2}=\left| A_{1}\right| ^{2}\exp
\left( -\frac{\left( x-X\right) ^{2}}{\left| a\right| ^{2}}\right) dx
\label{h2.3}
\end{equation}
\begin{equation}
j\left( x\right) =-\frac{i\hbar }{2m}\left( \psi ^{\ast }\frac{\partial \psi 
}{\partial x}-\frac{\partial \psi ^{\ast }}{\partial x}\psi \right) =\frac{k%
}{m}\left| A_{1}\right| ^{2}\exp \left( -\frac{\left( x-X\right) ^{2}}{%
\left| a\right| ^{2}}\right)  \label{h2.4}
\end{equation}
Then the mean momentum distribution (\ref{h1.4}) is described by the
relation 
\begin{equation}
w\left( \left\langle p\right\rangle \right) =\int \delta \left( \left\langle
p\right\rangle -k\right) \rho \left( x\right) dx=\left| A_{1}\right| ^{2}%
\frac{\sqrt{\pi }}{\left| a\right| }\delta \left( \left\langle
p\right\rangle -k\right)  \label{h2.5}
\end{equation}
Expressions (\ref{h2.5}) and (\ref{h2.2}) coincide in the limit $%
a\rightarrow \infty $. They are close, if $a\gg \hbar /k$. Thus,
distributions (\ref{h2.5}) and (\ref{h2.2}) are close for the wide wave
packets. In other cases the distributions\ (\ref{h2.5}) and (\ref{h2.2}) may
be close or not. Both interpretations (conventional and hydrodynamic) assume
that simultaneous distribution over position and momenta is impossible.
(Formally, such a distribution is possible, but in this distribution the
momentum $\mathbf{p}$ is a function of the position $\mathbf{x}$). But
reasons of this impossibility are different. The conventional interpretation
supposes that such a distribution is impossible, because the position
operator and the momentum operator do not commute. In the hydrodynamic
interpretation such a distribution is impossible, because the position and
the mean momentum are not independent quantities.

Let us consider the influence of the measuring device on the momentum
measurement in the stationary state described by the wave function $\psi $.
We measure the component $p_{1}$ of the momentum $\mathbf{p}$ in the
interval $\left\{ x,x+\Delta x\right\} $.

At first, we present the viewpoint of ACQP, which does not distinguish
between the $M$-measurement and $S$-measurement. We consider the momentum
measurement as an measurement under individual particle. We measure the
particle coordinate $x$ twice with the time interval $t_{\mathrm{m}}$
between them. For determination of the particle position, we use a photon of
frequency $\omega $ with the momentum $\hbar \omega /c$. At the collision
with the particle the photon transmits a portion of its momentum to the
particle. Indeterminacy $\Delta p$ of the momentum is connected with the
indeterminacy $\Delta x$ of the particle position by means of the
indeterminacy relation

\begin{equation}
\Delta x\Delta p\geq \hbar /2  \label{h2.7}
\end{equation}
Measurement of the first position of the particle is carried out by the
indeterminacy $\Delta x$. The momentum is determined by the relation (\ref
{a1.5}), and the indeterminacy $\delta p$ of the momentum measurement is a
function of $\Delta x$, defined by the relation

\begin{equation}
\delta p=\frac{m\Delta x}{t_{\mathrm{m}}}+\frac{\hbar }{\Delta x}
\label{h2.8}
\end{equation}
Minimum of (\ref{h2.8}) is attained at $\Delta x=\sqrt{m\hbar /t_{\mathrm{m}}%
}$. The minimal value of the momentum indeterminacy is given by the relation 
\begin{equation}
\delta p_{\min }=2\sqrt{\frac{m\hbar }{t_{\mathrm{m}}}}  \label{h2.9}
\end{equation}
Thus, the exact measurement of the momentum is possible only at $t_{\mathrm{m%
}}\rightarrow \infty $, when the measured value of the momentum cannot be
attributed to any value of the wave function. Impossibility of the exact
measurement of particle momentum is connected with influence of the
measuring device (in the given case with the photon influence).

Now we consider the momentum measurement from the viewpoint of MCQP, which
distinguishes between the $S$-measurement and the $M$-measurement.
Consideration of the $S$-measurement is useless, because the result of the $%
S $-measurement is random and irreproducible. Besides, the result of the $S$%
-measurement cannot be predicted by the statistical conception (quantum
mechanics).

The $M$-measurement is a set (statistical ensemble) of $N$ similar
independent $S$-measurements: $M_{1},$ $M_{2},...,M_{N}$ ($N\rightarrow
\infty $). Anyone of $S$-measurements is produced under a single particle.
All particles are prepared in the same way, and all $N$ particles form the
statistical ensemble $\mathcal{E}_{N}$, whose state is described by the wave
function $\psi $. Any $S$-measurement $M_{i}$ consists of two $S$%
-measurements $\left\{ F_{i},S_{i}\right\} $, where $F_{i}$ means the
measurement of the first position of the $i$th particle at the time $t$, and 
$S_{i}$ means the measurement of the position of the same particle at the
time $t+t_{\mathrm{m}}$. Let the $S$-measurements $F_{\alpha _{1}},F_{\alpha
_{2}},...F_{\alpha _{K}},$ $\left( K\gg 1\right) $ of the position $x$ give
the result in interval $L_{k}=\left( k\Delta x,\left( k+1\right) \Delta
x\right) $, where $L$ is some space interval. We select these particles with
numbers $\alpha _{1},\alpha _{2},...\alpha _{K}$ and form a new statistical
ensemble $\mathcal{E}_{K}^{\prime }$. The $K$ particles of the statistical
ensemble $\mathcal{E}_{K}^{\prime }$ have the position at the time $t$ in
the interval $L_{k}$. If $N\rightarrow \infty $ and $K\rightarrow \infty $,
both statistical ensembles $\mathcal{E}_{N}$ and $\mathcal{E}_{K}^{\prime }$
are dynamic systems, whose states are described respectively by the wave
functions $\psi $ and $\psi ^{\prime }$. The wave function $\psi ^{\prime }$
of the statistical ensemble $\mathcal{E}_{K}^{\prime }$ does not coincide,
in general, with the state $\psi $ of the statistical ensemble $\mathcal{E}%
_{N}$. This change of the ensemble state is mainly a result of the
selection, although the dynamic interaction with the measuring device
(photon) also contributes to the state of the statistical ensemble $\mathcal{%
E}_{K}^{\prime }$.

As a result of the produced selection the coordinate $x$ of all systems of
the statistical ensemble $\mathcal{E}_{K}^{\prime }$ is concentrated in the
interval $L_{k}=\left( k\Delta x,\left( k+1\right) \Delta x\right) $. Then $%
\left| \partial \rho ^{\prime }/\partial x\right| \geq \hbar \rho ^{\prime
}/\Delta x,$\ \ $\rho ^{\prime }=\left| \psi ^{\prime }\right| ^{2}$ and the
internal energy (\ref{h0.3a}) of the statistical ensemble $\mathcal{E}%
_{K}^{\prime }$ increases. After a lapse of time this increase of the
internal energy transforms into the $x$-component of the particle momentum.
Let us now produce the $S$-measurements $S_{\alpha _{1}},S_{\alpha
_{1}},...S_{\alpha _{K}}$ of coordinate $x$ of all $K$ particles of the
statistical ensemble $\mathcal{E}_{K}^{\prime }$ at the time $t+t_{\mathrm{m}%
}$. Using the relation (\ref{a1.5}), we obtain distribution $w_{t_{\mathrm{m}%
},\Delta x}\left( x,p_{x}\right) $ over the $x$-component of the momentum in
the vicinity of the point $x=k\Delta x$. Form of this distribution depends
on the measurement time $t_{\mathrm{m}}$ and on the influence of the
measuring device. The distribution takes into account also the stochastic
component of the particle $\mathcal{S}_{\mathrm{st}}$ motion. If the numbers 
$N$ and $K$ of elements of the statistical ensembles $\mathcal{E}_{N}$ and $%
\mathcal{E}_{K}^{\prime }$ tend to $\infty $, the obtained distribution $%
w_{t_{\mathrm{m}},\Delta x}\left( x,p_{x}\right) $ appears to be
reproducible. It is reproduced at repeated $M$-measurements of the $x$%
-component of the momentum.

The obtained distribution $w_{t_{\mathrm{m}},\Delta x}\left( x,p_{x}\right) $
over momenta depends on $x=k\Delta x$, on $t_{\mathrm{m}}$ and on $\Delta x$
which are chosen arbitrarily. We can consider the limit 
\begin{equation}
w_{\Delta x}\left( x,p_{x}\right) =\lim_{t_{\mathrm{m}}\rightarrow 0}w_{t_{%
\mathrm{m}},\Delta x}\left( x,p_{x}\right)   \label{h2.11}
\end{equation}
We cannot be sure that such a limit exists, because we know nothing about
the stochastic evolution component $\mathcal{C}_{\mathrm{st}}$ of the
particle $\mathcal{S}_{\mathrm{st}}$ motion. As far as all particles of the
statistical ensemble $\mathcal{E}_{K}^{\prime }$ are independent, we should
expect, that influence of the quantum stochasticity and influence of the
measuring device on the mean value $\left\langle p_{x}\right\rangle $ of the
momentum $x$-component is compensated. One should expect that $w_{\Delta
x}\left( x,p_{x}\right) $ depends only on $x=k\Delta x$, but not on $\Delta x
$, i.e. $w_{\Delta x}\left( x,p_{x}\right) =w\left( x,p_{x}\right) $. Then
the mean value of the particle momentum $\left\langle p_{x}\left( x\right)
\right\rangle $ on the interval $L_{k}=\left( k\Delta x,\left( k+1\right)
\Delta x\right) $ 
\begin{equation}
\left\langle p_{x}\left( x\right) \right\rangle =\int p_{x}w\left(
x,p_{x}\right) dp_{x}  \label{h2.10}
\end{equation}
exists and coincides with the value of $mv_{x}$, determined by the second
relation of (\ref{h1.4}). The measured value $\left\langle p_{x}\left(
x\right) \right\rangle $ of the mean momentum may be attributed to the state 
$\psi $ (but not to $\psi ^{\prime }$), because the time of the measurement
can be made to be very short. The state $\psi ^{\prime }$ may be considered
to be an intermediate state, which appears in the process of the $M$%
-measurement. The quantum mechanics formalism predicts the value of (\ref
{h2.10}) and attributes it to the wave function $\psi $. However, the QM\
formalism cannot predict the form of the distribution (\ref{h2.11}), as well
as the gas dynamics can describe the gas velocity (the mean molecular
velocity), but the gas dynamics formalism does not describe the Maxwell
distribution and its evolution. In this sense the parameters of the
distribution (\ref{h2.11}) may be considered to be hidden parameters of
ACQP. They are also hidden parameters of MCQP.

In a sense the momentum distribution (\ref{h2.11}) is an analog of the
Maxwell momentum distribution (\ref{a1.3}) in the kinetic gas theory. Is it
possible to measure this momentum distribution experimentally? Is it
possible to eliminate influence of the measuring device on the measurement
of the momentum distribution? It is an interesting question, which is open
now.

\section{Concluding remarks}

At first sight the conventional interpretation seems to be more informative,
than the hydrodynamic interpretation, because the ACQP predicts the momentum
distribution, whereas MCQP cannot. We show in this paper that the momentum
distribution (\ref{h0.3}) is not a momentum distribution in reality. It may
be considered to be the Fourier component module distribution, or even the
mean momentum distribution, but it is not the momentum distribution. If it
is so, the hydrodynamic interpretation, which cannot predict the momentum
distribution, appears to be as informative as the conventional
interpretation of quantum mechanics.

Besides it was shown that the wave function describes the state of the
statistical ensemble of stochastic particles $\mathcal{E}\left[ \mathcal{S}_{%
\mathrm{st}}\right] $, or the state of the statistical average particle $%
\left\langle \mathcal{S}_{\mathrm{st}}\right\rangle $, but not the state of
an individual stochastic particle $\mathcal{S}_{\mathrm{st}}$. It is a
serious argument against the conventional interpretation of quantum
mechanics.

Thus, we have shown that the quantum system $\mathcal{S}_{\mathrm{q}}$ is
the statistical average system $\left\langle \mathcal{S}_{\mathrm{st}%
}\right\rangle $, but not an individual system $\mathcal{S}_{\mathrm{st}}$.
The quantum mechanics formalism predicts only results of the $M$%
-measurements. The $S$-measurements are important, because any $M$%
-measurement is an  ensemble of independent $S$-measurements.

It has been shown that some of original basic statements (\ref{h0.1}), which
are necessary for the proof of the von Neumann's theorem on hidden variables
cannot be tested experimentally, because the measured values cannot be
attributed to a definite state (wave function). All these arguments attest
in favour of hydrodynamic interpretation, which can be freely expanded to
the case of relativistic phenomena.

\end{document}